# A new family of disorder-free Rare-Earth-based kagomé lattice magnets: structure and magnetic characterizations of $RE_3BWO_9$ (RE=Pr, Nd, Gd-Ho) Boratotungstates

Malik Ashtar,[†] Jinjin Guo,[‡] Zongtang Wan,[†] Yongqiang Wang,[‡] Gaoshang Gong,[‡] Yong Liu,[§] Yuling Su,[*,‡] Zhaoming Tian[*,†]

[†]School of Physics and Wuhan National High Magnetic Field centre, Huazhong University of Science and Technology, Wuhan 430074, Peoples R China.

[‡] Zhengzhou Univ Light Ind, Sch Phys & Elect Engn, Zhengzhou 450002, Henan, Peoples R China

[§] School of Physics, Wuhan University, Wuhan 430072, P.R. China

## ABSTRACT:

Exploration of rare-earth (RE)-based Kagomé lattice magnets with spin-orbital entangled $j_{eff}$=1/2 moments will provide new platform for investigating the exotic magnetic phases. Here, we report a new family of $RE_3BWO_9$ (RE=Pr,Nd,Gd-Ho) boratotungstates with magnetic $RE^{3+}$ ions arranged on Kagomè lattice, and perform its structure and magnetic characterizations. This serial compounds crystallize in hexagonal coordinated structure with space group $P6_3$ (No.173), where magnetic $RE^{3+}$ ions have distorted Kagomé lattice connections within the *ab* plane and stacked in a AB-type fashion along *c* axis. The interlayer RE-RE separation is comparable with that of intralayer distance, forming 3-dimensional (3D) exchange coupled magnetic framework of $RE^{3+}$ ions. The magnetic susceptibility data of $RE_3BWO_9$ (RE=Pr, Nd, Gd-Ho) reveal dominant antiferromagnetic interactions between magnetic $RE^{3+}$ ions, but without visible magnetic ordering down to 2 K. The magnetization analyses for different $RE^{3+}$ ions show diverse anisotropic behaviors, make $RE_3BWO_9$ as an appealing Kagomé-lattice antiferromagnet to explore exotic magnetic phases.

## ■ INTRODUCTION

Geometrically frustrated magnets are an active research field in area of condensed-matter physics, which can provide ideal platforms for exploring exotic magnetic ground states, such as quantum spin-liquid (QSL) state, chiral spin ordered state, spin ice and spin nematic or multipolar state, etc. In such systems, the presence of antiferromagnetic (AFM) exchange interactions on crystal lattice geometry with corner-or edge-sharing triangle or tetrahedra can exhibit geometric magnetic frustration, supporting the exotic forms of magnetism. The Kagomé lattice, composed of six corner shared triangles surrounding a hexagon, are quite attractive for search of QSL state due to their strong geometrical frustration, low coordination and weak second-neighbour coupling.[1-3] In past several decades, the prototypical examples of Kagomé antiferromagnet have been heavily studied based on 3d transition metal systems, such as Herbertsmithite $ZnCu_3(OH)_6Cl_2$[4], Vesignieite $BaCu_3V_2O_8(OH)$,[5] fluoride $NaBa_2Mn_3F_{11}$,[6] and $BaNi_3(OH)_2(VO_4)_2$.[7] Besides the fascinating physics associated QSL ground state, the presence of fractionalized excitation in the QSL materials can also be useful for future design of the topological computations.[8] However, the experimental identification of QSL state still remains to ongoing debate. Most of existing QSL candidates suffer from the structural disorder effect, as one of most heavily studied Kagomé lattice $ZnCu_3(OH)_6Cl_2$, few percent of $Cu^{2+}/Zn^{2+}$ induces randomness of magnetic coupling, cause the difficulty in understanding its ground state.[9] Additionally, the roadblocks in the identification of exotic phases in these materials are strongly affected by the issues of site disorder, weak magnetocrystalline anisotropy, weak magnetic exchange coupling and mixture with spin-glass phase.[10] Thus, discovery of new materials free of chemical site-mixing occupancy to avoid the influence of exchange disorder on its' magnetic behaviors, are essential to clarify the intrinsic magnetic behaviors.

To pursuit experimental realization of exotic magnetic phases, frustrated magnets incorporating heavy-element-based magnetic ions, such as 5d transition metal or 4f rare-earth (RE) ions, have recently been intensive investigated due to the enhanced quantum fluctuations for the spin-orbit entangled $j_{eff}$=1/2 moments.[3] Furthermore, the diverse spin-types and various spin anisotropies of different $RE^{3+}$, multiple interplay among spin-orbital coupling (SOC), exchange and dipolar interactions in combination with geometric lattice frustration will lead to a wider variety of exotic magnetic phenomena.[11-13] In metallic compounds with RE ions located on kagomé lattice, two serial $RE_3Ru_4Al_{12}$ and REPtPb compounds are recently well studied,[14,15] which are attractive to explore exotic physical phenomena related to the interplay between itinerant electrons and noncollinear magnetic structures. On other hand, insulating RE-based oxides are also interesting to unveil the pure Kagomé physics fully dictated by topological magnetic frustrations of RE ions, excluding the influence of conductive electrons. In this respect, the related materials are still rare, only two families of RE-based Kagomé lattice compounds $RE_3Sb_3M_2O_{14}$ (RE = Pr, Nd, Sm-Yb, M = Zn, Mg)[16-18] and $RE_3Ga_5SiO_{14}$(RE = Pr, Nd)[19] are magnetically studied, to the best our knowledge. The $RE_3Sb_3M_2O_{14}$ (RE = Pr,Nd,Sm-Yb, M = Zn,Mg) antiferromagnets are proposed to display interesting magnetic ground states including Kagomé spin ice, dipolar spin-order and scalar spin-chirality.[16] The discovery of $RE_3Ga_5SiO_{14}$ provide experimental form of QSL candidate with local spin correlation at millikelvin temperature scale.[20] Therefore, exploring new RE-based magnets with Kagomé lattice are highly desirable to uncover long-sought QSL state and other exotic magnetic phenomena.

Here, we report the synthesis, structure and magnetic properties of new RE-based frustrated antiferromagnets, $RE_3BWO_9$ (La, Pr, Nd, Gd-Ho) boratotungstates, crystallized into the hexagonal structure with $P6_3$ space group.[21] The $RE^{3+}$ ions carrying magnetic spins are arranged on a distorted

Kagomé lattice within the *ab* plane and stacked in an alternating AB type fashion along *c* axis. For this serial compounds, both magnetic $RE^{3+}$ ions and nonmagnetic $W^{6+}/B^{3+}$ cations form the ordered state where chemically antisite disorder are well avoided thanks to their large difference of ionic radii and coordination number in these boratotungstates. The magnetic susceptibilities reveal all members of series have a dominant AFM spin interaction with no ordering down to 2 K.

## ■ EXPERIMENTAL SECTION

**Material Synthesis.** High-quality $RE_3BWO_9$ (La, Pr, Nd, Gd-Ho) polycrystalline samples were synthesized by a solid-state reaction method using $H_3BO_3$, $WO_3$, and RE oxides as starting materials. To ensure stoichiometry, the RE oxides ($La_2O_3$, $Pr_2O_3$, $Nd_2O_3$, $G_2O_3$, $Tb_3O_7$, $Dy_2O_3$, and $Ho_2O_3$) were dried at 900 °C overnight before using. The appropriate mixtures of starting materials were thoroughly mixed and sintered in temperature range 1000 °C to 1200 °C for several days with intermediate re-grinding.

**Characterization.** The powder X-ray diffraction (XRD) data of $RE_3BWO_9$ (La, Pr, Nd, Gd-Ho) compounds were collected at room temperature using a PAN analytical X'Pert Pro MPD diffractometer with Cu Kα radiation= 1.5418 Å. Then, the structural analyses were carried out by the Rietveld refinements using Material studio software. The magnetic susceptibilities from 2 K – 300 K and field-dependent isothermal magnetization at different temperature up to 7 T were measured using a SQUID magnetometer (MPMS, Quantum Design). The isothermal field-dependent magnetization measurements from 0 T to 14 T were performed using vibrating sample magnetometer (VSM) equipped with the physical properties measurement system (PPMS, Quantum Design).

## ■ RESULTS AND DISCUSSION

**Structural description.** Room temperature powder XRD patterns for representative boratotungstates $RE_3BWO_9$ (La, Nd, Ho) compounds are displayed in Figure 1a. The observed XRD spectra can be indexed into hexagonal coordinated structure with space group $P6_3$ (No.173), without other detected impurity phases. To obtain detailed structural information, the crystal structure is refined using $La_3BWO_9$ as a starting model by Rietveld method, the lattice parameters including lattice constants and atomic positions of $RE_3BWO_9$ (La, Pr,Nd, Gd-Ho) boratotungstates are consistent with the prior reports,[21] as listed in Table 1. The refinement yields a good fit between the observed and simulated profiles with refinement reliability factors $R_p$ (2.79- 4.86) and $R_{wp}$ (4.05 - 6.52). With the variation of $RE^{3+}$ ionic radii, the lattice constants (*a, b, c*) and unit-cell volume follow a linear dependence, this can be related to the Lanthanide contraction effect (Figure 1b,c). The selected interatomic bonds and angles pertinent to RE magnetic ions for their intralayer and interlayer separations are summarized in Table 2.

Since the serial RE$_3$BWO$_9$ (RE = Pr, Nd, Gd-Ho) compounds are isostructural, the crystal structure of Pr$_3$BWO$_9$ as a representative is schematically shown in Figure 2a. The unit cell of hexagonal Pr$_3$BWO$_9$ consists of one unique type of Pr atom (Wyckoff site 6c), one B atom (Wyckoff site 2a), one W atom (Wyckoff site 2b) and three O atoms (Wyckoff site 6c). The individual coordination of PrO$_8$, WO$_6$, and BO$_3$ polyhedra are presented in Figure 2c-e. The Pr is positioned in a distorted dodecahedron site coordinated by 8 oxygens with three different Pr–O distances 2.651(1) Å, 2.542(1) Å and 2.333(2) Å. The W atom is centered in the distorted octahedral geometry of 6 surrounding oxygen atoms with W–O$_2$ and W–O$_3$ bond lengths of 1.972(4) Å and 1.889(2) Å, respectively, the O–W–O angles fall in the range of 81.2(5)° to 159.4(3)°. Boron is located at the center in BO$_3$ trigonal coordination. In terms of structural linkages, the WO$_6$ octahedra and BO$_3$ trigons share edges with three PrO$_8$ polyhedra, connecting PrO$_8$ polyhedrons as the bridging ligand. Along *c*-axis, the PrO$_8$ polyhedra share edges with the neighboring PrO$_8$ polyhedra between different layers (Figure 2b). The tungsten ions reside nearly at the center of the distorted hexagon formed by RE cations in Pr$_3$BWO$_9$ structure.

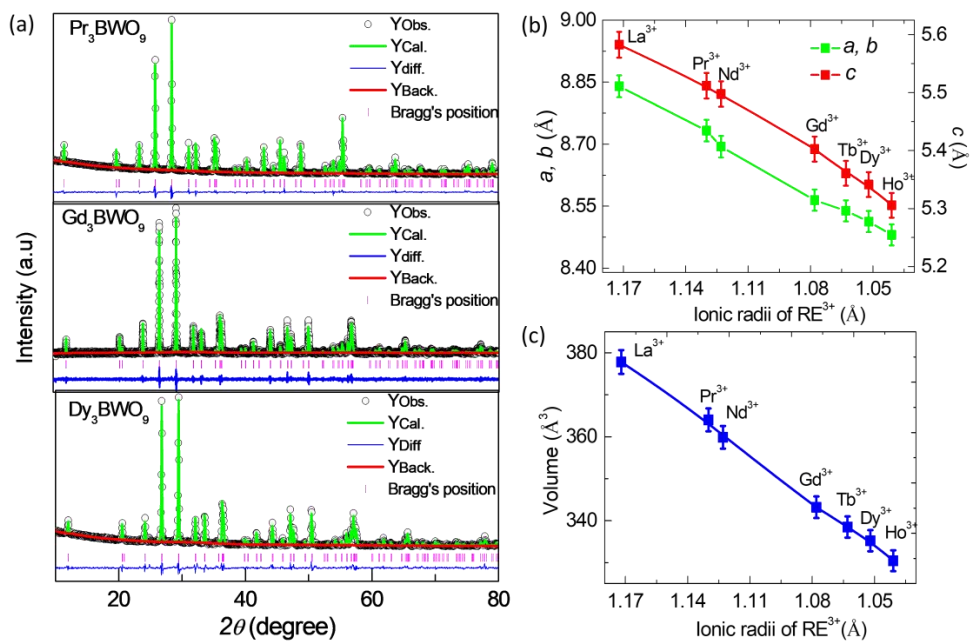

Fig. 1. (a) The experimental and refined XRD patterns of RE$_3$BWO$_9$ (La, Gd, Dy) compounds. The open circles is experimental data, green line is calculated data, the blue, red lines show the difference and background, and pink marks indicate brags reflections. (b) The variation of lattice parameters as a function of RE ionic radius. (c) The variation of unit-cell volume as a function of ionic radius.

Table 1. Refined structural parameters of RE$_3$BWO$_9$ from room-temperature XRD patterns.

| Crystallographic Parameters | La | Pr | Nd | Gd | Tb | Dy | Ho |
|---|---|---|---|---|---|---|---|
| crystal system | Hexagonal | Hexagonal | Hexagonal | Hexagonal | Hexagonal | Hexagonal | Hexagonal |
| space group | P6$_3$ (173) | P6$_3$ (173) | P6$_3$ (173) | P6$_3$ (173) | P6$_3$ (173) | P6$_3$ (173) | P6$_3$ (173) |
| a(Å) | 8.8399 | 8.7327 | 8.6939 | 8.5646 | 8.5386 | 8.5128 | 8.4805 |
| c(Å) | 5.5827 | 5.5119 | 5.4976 | 5.4026 | 5.3608 | 5.3414 | 5.3056 |
| V(Å$^3$) | 377.81 | 364.02 | 359.86 | 343.20 | 338.48 | 335.22 | 330.45 |
| α , β | 90 | 90 | 90 | 90 | 90 | 90 | 90 |
| γ | 120 | 120 | 120 | 120 | 120 | 120 | 120 |
| RE (6c) | (x, y, z) | (x, y, z) | (x, y, z) | (x, y, z) | (x, y, z) | (x, y, z) | (x, y, z) |
| x | 0.0824 | 0.0818 | 0.0804 | 0.0788 | 0.0794 | 0.0784 | 0.0757 |
| y | 0.7232 | 0.7247 | 0.7241 | 0.7229 | 0.7234 | 0.7225 | 0.7264 |
| z | 0.7170 | 0.7016 | 0.7036 | 0.6989 | 0.7005 | 0.6990 | 0.7082 |
| U (Å) | 0.0039 | 0.0031 | 0.0510 | 0.0028 | 0.0116 | 0.0083 | 0.0328 |
| B (2a) | (0, 0, z) | (0, 0, z) | (0, 0, z) | (0, 0, z) | (0, 0, z) | (0, 0, z) | (0, 0, z) |
| z | 0.3831 | 0.3646 | 0.3689 | 0.3688 | 0.3716 | 0.3734 | 0.3779 |
| U (Å) | 0.6021 | 0.2295 | 0.6188 | 0.6565 | 0.5783 | 0.7274 | 0.6706 |
| W (2b) | (1/3, 2/3, z) | (1/3, 2/3, z) | (1/3, 2/3, z) | (1/3, 1/3, z) | (1/3, 2/3, z) | (1/3, 2/3, z) | (1/3, 2/3, z) |
| z | 0.2542 | 0.2444 | 0.2471 | 0.2417 | 0.2437 | 0.2419 | 0.2485 |
| U (Å) | 0.0530 | 0.0331 | 0.0294 | 0.0724 | 0.0330 | 0.0149 | 0.2929 |
| O1 (6c) | (x, y, z) | (x, y, z) | (x, y, z) | (x, y, z) | (x, y, z) | (x, y, z) | (x, y, z) |
| x | 0.0487 | 0.0469 | 0.0481 | 0.0543 | 0.0565 | 0.0595 | 0.0540 |
| y | 0.8749 | 0.8725 | 0.8725 | 0.8750 | 0.8755 | 0.8775 | 0.8765 |
| z | 0.3762 | 0.3568 | 0.3606 | 0.3602 | 0.3627 | 0.3645 | 0.3712 |
| U (Å) | 0.0158 | 0.0339 | 0.0385 | 0.0836 | 0.0431 | 0.0201 | 0.4543 |
| O2 (6c) | (x, y, z) | (x, y, z) | (x, y, z) | (x, y, z) | (x, y, z) | (x, y, z) | (x, y, z) |
| x | 0.1391 | 0.1369 | 0.1352 | 0.1321 | 0.1308 | 0.1299 | 0.1351 |
| y | 0.4731 | 0.5210 | 0.5198 | 0.5190 | 0.5175 | 0.5180 | 0.5119 |
| z | 0.5226 | 0.4671 | 0.4708 | 0.4690 | 0.4743 | 0.4728 | 0.4676 |
| U (Å) | 0.0392 | 0.0471 | 0.6419 | 0.2161 | 0.0408 | 0.0171 | 0.03740 |
| O3 (6c) | (x, y, z) | (x, y, z) | (x, y, z) | (x, y, z) | (x, y, z) | (x, y, z) | (x, y, z) |
| x | 0.2003 | 0.2042 | 0.2060 | 0.2038 | 0.2020 | 0.2023 | 0.2051 |
| y | 0.4714 | 0.4684 | 0.4663 | 0.4634 | 0.4622 | 0.4613 | 0.47383 |
| z | 0.0492 | 0.0415 | 0.0445 | 0.0345 | 0.0352 | 0.0326 | 0.0407 |
| U (Å) | 0.0259 | 0.06382 | 0.0239 | 0.0537 | 0.2472 | 0.1828 | 0.1469 |
| $R_p$ | 3.59 | 4.17 | 3.96 | 4.86 | 3.13 | 2.79 | 4.37 |
| $R_{wp}$ | 5.93 | 5.98 | 5.62 | 6.52 | 4.74 | 4.05 | 6.11 |

Table 2. Selected bond distances (Å), bond angles, interplane and intraplane RE–RE distances of serial $RE_3BWO_9$ compounds.

| RE | La | Pr | Nd | Gd | Tb | Dy | Ho |
|---|---|---|---|---|---|---|---|
| $REO_9$ | | | | | | | |
| $RE–O_1$ | 2.7196 | 2.6508 | 2.5199 | 2.4434 | 2.4237 | 2.3995 | 2.4017 |
| $RE–O_2$ | 2.6801 | 2.6215 | 2.4888 | 2.4376 | 2.3515 | 2.3339 | 2.3421 |
| $RE–O_3$ | 2.4044 | 2.3770 | 2.3634 | 2.3085 | 2.2942 | 2.2822 | 2.2782 |
| Interplane RE-RE Å | 4.012 | 3.9503 | 3.9337 | 3.8712 | 3.8499 | 3.8387 | 3.8265 |
| Intraplane RE-RE Å | 4.3408 | 4.3102 | 4.3067 | 4.2550 | 4.2382 | 4.2311 | 4.2248 |
| $O_1–RE–O_2$ (°) | 95.112 | 95.679 | 95.675 | 96.062 | 96.712 | 96.908 | 95.778 |
| $O_1–RE–O_3$ (°) | 136.241 | 138.239 | 138.214 | 136.045 | 136.017 | 134.455 | 133.846 |
| $O_2–RE–O_3$ (°) | 77.494 | 78.80 | 79.370 | 79.064 | 79.241 | 79.836 | 79.455 |
| $WO_6$ | | | | | | | |
| $W–O_2$ | 1.9692 | 1.9726 | 1.9773 | 1.9746 | 1.9843 | 1.9830 | 1.9003 |
| $W–O_3$ | 1.9082 | 1.8888 | 1.8903 | 1.8927 | 1.8966 | 1.8972 | 1.8518 |
| $O_2–W–O_2$ (°) | 85.524 | 85.395 | 85.408 | 85.396 | 85.294 | 85.347 | 86.488 |
| $O_2–W–O_3$ | 158.295 | 159.436 | 160.462 | 160.764 | 160.338 | 160.820 | 1161.061 |
| $O_3–W–O_3$ | 87.746 | 88.524 | 88.795 | 88.602 | 88.802 | 88.849 | 88.176 |
| $BO_3$ | | | | | | | |
| $B–O_1$ | 1.3732 | 1.3655 | 1.3668 | 1.3645 | 1.3698 | 1.3689 | 1.3255 |
| $O_1–B–O_1$ (°) | 119.922 | 119.900 | 119.891 | 119.884 | 119.878 | 119.881 | 119.929 |

For $RE_3BWO_9$, the RE–O distances decrease as changing RE from La to Ho, whereas the W–O and B–O bond distances have small variation up to $Dy_3BWO_9$. In $WO_6$ polyhedron, the shortest W–O distance decreases while longest W–O distance increases slightly as RE going from La to Dy, result in the enhanced distortion of polyhedron. The variation of B–O and W–O distances are also observable, reveals that the lattice variation is induced by increasing distortions of $REO_8$ and $WO_6$ polyhedron in the $RE_3BWO_9$ series. Considering that the large difference of ionic radii and coordination numbers of RE and B/W cations, the site-mixing disorder can be avoided. Namely, the RE, W and B cations in $RE_3BWO_9$ prefer to form fully ordered arrangements, correspond to the nearly 100% occupation of $RE^{3+}$ ions on Kagomé lattice.

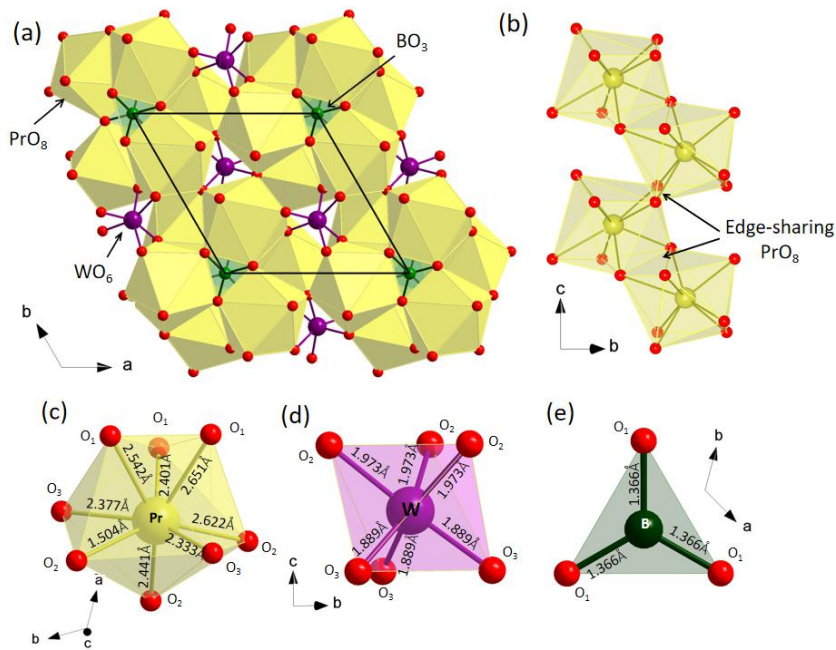

Fig. 2. (a) Top view of crystal structure of $Pr_3BWO_9$ along $c$ direction. (b) The edge-shared $PrO_8$ polyhedron within $bc$ plane. (c) The coordination environment for the $PrO_8$, $WO_6$ and $BO_3$ building blocks.

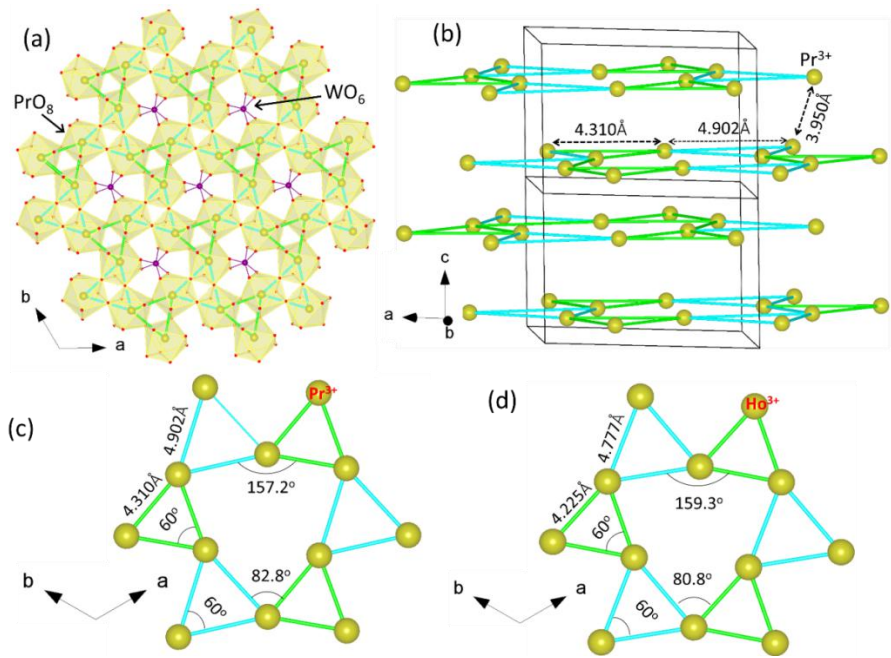

Fig. 3. (a) The top view of $PrO_8$ polyhedrons and its connections of $Pr_3BWO_9$ in the $ab$ plane, the cyan and green solid lines show the Kagomé networks. (b) The well separated Kagomé planes are staked in alternative AB fashion along $c$ axis of $Pr_3BWO_9$. The comparisons of Kagomé lattice for (c) $Pr^{3+}$ ions and (d) $Ho^{3+}$ ions.

In Pr$_3$BWO$_9$, the topological arrangement of magnetic Pr$^{3+}$ ions can be reflected by the connections of PrO$_8$ polyhedrons, as outlined in Figure 3a,b, where Pr triangles form a corner-sharing distorted Kagomé network in the ab plane. The WO$_6$ polyhedra reside at the center of Kagomé hexagon and are interlinked with surrounding three PrO$_8$ by edge-sharing. Along c direction, the Pr$^{3+}$ based-Kagomé lattices are stacked in a AB-type fashion with equal distance. One of important factors affecting magnetic interactions are the Pr$^{3+}$-Pr$^{3+}$ distances, the interlayer separations between Pr$^{3+}$ ions are 3.950(1) Å, slightly smaller than that of intralayer distances 4.310(2) Å and 4.901(2) Å, as represented by the green and cyan equilateral triangular side lengths in Figure 3b, respectively. Here, the similar intralayer and interlayer Pr-Pr distances enforce the three dimensional (3D) magnetic interactions as the situation in RE-based honeycomb SrRE$_2$O$_4$ [22] and Pyrochlore systems.[23,24] This is different from the layered fluoride NaBa$_2$Mn$_3$F$_{11}$,[6] where magnetic Mn$^{2+}$ ions form two-dimensional (2D) distorted Kagomé lattice. As the RE$^{3+}$ ions change from Pr$^{3+}$ to Ho$^{3+}$, it leads to slightly different distortion of Kagomé lattice of magnetic ions, reflected by its internal angles of hexagon from 157.213° to 159.245° and 82.787° to 80.755°, as shown in Figure 3c,d. Here, the free of structural anti-site disorder is critical, rendering RE$_3$BWO$_9$ as a clean system where magnetic behavior originate from the intrinsic ordered Kagomé lattice without magnetic exchange randomness.

Table 3. Magnetic parameters obtained from the fitting of χ(T) by the Curie–Weiss law, Curie–Weiss temperatures ($\theta_{CW}$) and effective magnetic moments ($\mu_{eff}$), and the effective moment ($\mu_{fi}$) expected for free ions calculated by $g[J(J+1)]^{1/2}$ for RE$_3$BWO$_9$ (RE = Pr, Nd, Gd–Ho) compounds.

| RE | High T fit (K) | $\theta_{CW}$ (K) | $\mu_{eff}$ ($\mu_B$) | Low T fit (K) | $\theta_{CW}$ (K) | $\mu_{eff}$ ($\mu_B$) | $\mu_{fi}$ ($\mu_B$) |
|---|---|---|---|---|---|---|---|
| Pr | 100-300 | -40.30(1) | 3.55(5) | 8-25 | -6.88(3) | 3.09(3) | 3.58 |
| Nd | 100-300 | -49.22(1) | 3.71(1) | 8-25 | -2.20(5) | 2.85(6) | 3.62 |
| Gd | 100-300 | -3.64(5) | 8.65(5) | 8-25 | -0.89(2) | 8.52(1) | 7.94 |
| Tb | 100-300 | -3.24(2) | 10.03(2) | 8-25 | -0.79(4) | 9.61(3) | 9.72 |
| Dy | 100-300 | -5.81(6) | 10.86(3) | 8-25 | -0.82(2) | 10.49(3) | 10.63 |
| Ho | 100-300 | -9.28(1) | 10.95(5) | 8-25 | -1.14(2) | 10.38(1) | 10.60 |

**Magnetic properties.** For the serial RE$_3$BWO$_9$ (La,Pr,Nd,Gd-Ho) compounds, temperature (T) dependence of magnetic susceptibility χ(T) was measured from 2 K to 300 K with magnetic field H = 1000 Oe. The inverse susceptibilities are analyzed by Curie-Weiss law $1/\chi = (1/C)(T-\theta_{CW})$, the extracted $\theta_{CW}$ and $\mu_{eff}$ from susceptibility data are listed in Table 3, including comparison to the effective moments of free RE$^{3+}$ ions. Considering that magnetic contribution from the excited states of crystal electric field (CEF) splitting levels can change at different temperatures, the Curie-Weiss fitting of $\chi(T)$ were performed at two different temperature regimes 300 K - 100 K and 8 K - 25 K, respectively. Low temperature fitting can better reflect magnetic exchange interaction of effective spins on Kagomé lattice, where most electrons occupy at the CEF ground state. The $\mu_{eff}$ is calculated by $\mu_{eff} = (3k_B C/N_A)^{1/2}$ equation, where $k_B$ is Boltzmann constant, and $N_A$ is Avogadro's constant. The isothermal magnetizations $M(H)$ at different temperatures are analysed by the Brillouin function $B_J(x) = \frac{2J_{eff}+1}{2J_{eff}} coth \frac{2J_{eff}+1}{2J_{eff}} x - \frac{1}{2J_{eff}} coth \frac{x}{2J_{eff}}$, where $x = g_J \mu_B J_{eff} \mu_0 H / k_B T$, $g_J$ is the Lande's factor, $J_{eff}$ is the fitted effective angular momentum and $\mu_B$ is Bohr magneton, respectively.

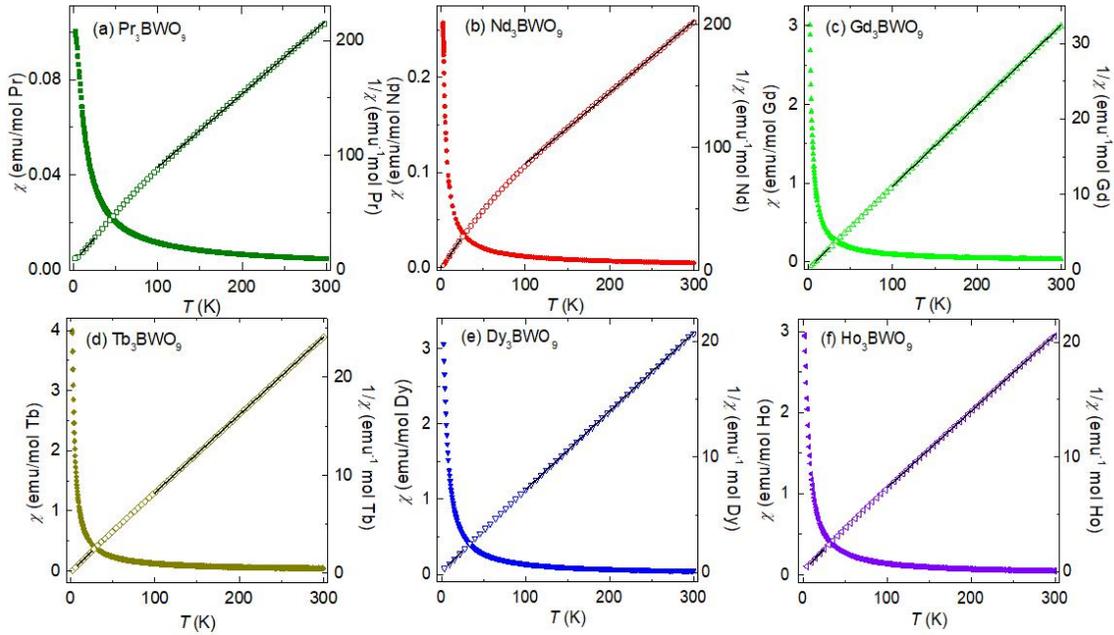

Fig. 4. Temperature-dependent magnetic susceptibility and inverse susceptibility measured at 0.1 T for RE$_3$BWO$_9$ (RE=Pr,Nd,Gd,Tb,Dy,Ho), respectively. The solid black lines show the Curie–Weiss fitting.

**Pr$_3$BWO$_9$.** The magnetic susceptibility $\chi(T)$ of Pr$_3$BWO$_9$ show no sign of magnetic transitions down to $T$ = 2 K (Figure 4a). At high temperatures ($T$ > 100 K), the inverse magnetic susceptibility $\chi^{-1}(T)$ exhibits typical Curie-Weiss behavior, giving rise to $\theta_{CW}$ = -40.30 K and $\mu_{eff}$ = 3.55 $\mu_B$/Pr$^{3+}$ close to the free ion value $g_J[J(J+1)]^{1/2}$ = 3.58 $\mu_B$/Pr$^{3+}$. Further decreasing temperatures, $\chi^{-1}(T)$ deviates from the Curie-Weiss law followed by another slope change below 40 K, reveals the variation of magnetic interaction and effective moment in this temperature regime. The low-temperature (8 K - 25 K) fit provides $\mu_{eff}$ = 3.09 $\mu_B$/Pr$^{3+}$ and $\theta_{CW}$ = -6.88 K, respectively. The negative $\theta_{CW}$ reveals a dominant antiferromagnetic (AFM) interaction between Pr$^{3+}$ spins persistent to the lowest measured temperatures. The variation of $\theta_{CW}$ and reduced $\mu_{eff}$ can be induced by the thermal population of electrons occupied on the CEF levels within the crystal lattice,[25,26] where most electrons are apt to occupy the CEF ground level as decreased temperatures.

The isothermal $M(H)$ curves with field up to 14 T are collected at different temperatures, as shown in Figure 5a. At 2 K, the magnetization exhibits a nonlinear field response with maxima value ~1.21 $\mu_B$/Pr$^{3+}$ at 14 T, this value is far from saturated theoretical moments $M_S = g_JJ\mu_B$ = 3.2 $\mu_B$/Pr$^{3+}$ for free Pr$^{3+}$ ions, but similar to the observed value in other Pr$^{3+}$ based frustrated magnets, e.g. Pr$_3$M$_2$Sb$_3$O$_{14}$ (M=Mg, Zn),[17,18] PrMAl$_{11}$O$_{19}$ (M=Mg, Zn),[27,28] etc. The $M(H)$ data were fitted with $M = g_JJ_{eff}\mu_BB_J(x)$ equation, $x$ is written as $x = g_J\mu_BJ_{eff}\mu_0H/k_BT$ for the measured temperature regions. The high-temperature ($T > 5$ K) $M(H)$ data can be well fitted to Brillouin function which reveals the PM nature of the system. The deviation from Brillouin function fitting for $M(H)$ curves at 2 K support the development of AFM exchange interactions between Pr$^{3+}$ moments.

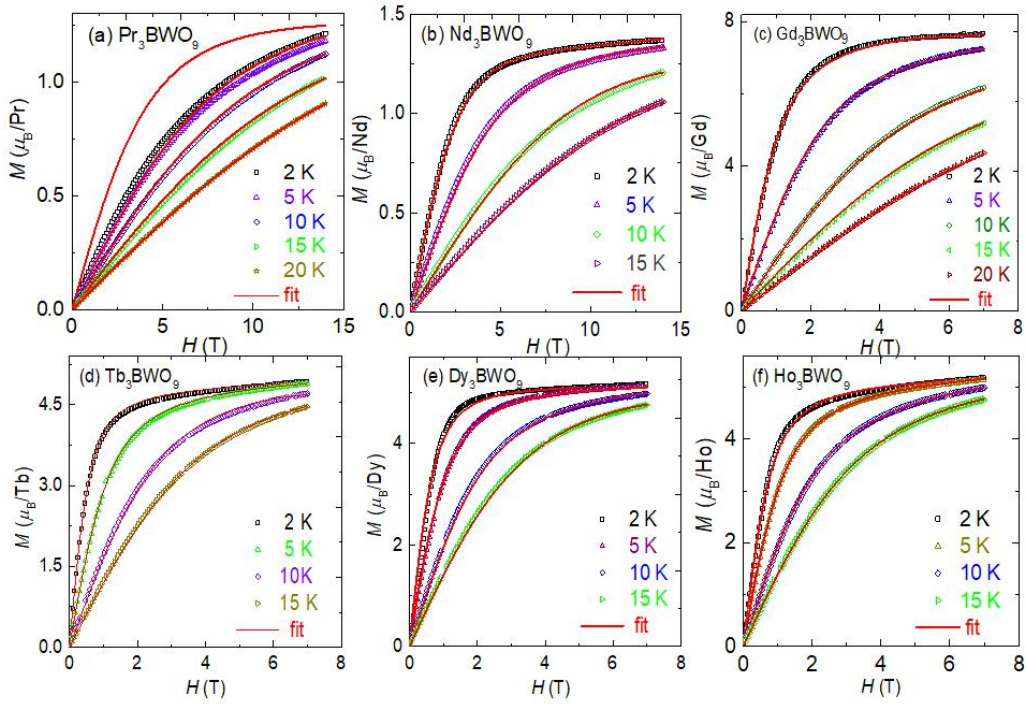

Fig. 5. The isothermal field-dependent magnetization curves at different temperatures for RE$_3$BWO$_9$ (RE=Pr,Nd,Gd,Tb,Dy,Ho), respectively. The solid red lines show the Brillouin function fits.

**Nd$_3$BWO$_9$.** Figure 4b shows the $\chi(T)$ of Nd$_3$BWO$_9$, which show no magnetic ordering down to 2 K. The $\chi^{-1}(T)$ of Nd$_3$BWO$_9$ is well fitted by the Curie-Wiess law above 100 K, yielding $\theta_{CW}$ = -49.2 K. The experimental obtained $\mu_{eff}$ = 3.71 $\mu_B$/Nd$^{3+}$ consistent with the expected value for free Nd$^{3+}$ ions $g_J[J(J+1)]^{1/2}$ = 3.62 $\mu_B$/Nd$^{3+}$. As decreased temperatures, magnetic contributions from the excited levels decrease due to the variable thermal populations on CEF levels, lead to the change in the slope of inverse susceptibility. The linear fit of $\chi^{-1}$(T) gives $\theta_{CW}$ = -2.20 K and reduced moment $\mu_{eff}$ = 2.85$\mu_B$/Nd$^{3+}$. This observed reduced $M_S$ can be due to the CEF effect. Since the Nd$^{3+}$ ($^4f_3$, $^4I_{9/2}$) as Kramer ion splits into five degenerate Kramers' doublets under the CEF effect, which can affect the spin correlations and magnetic moment.[18,27] The $M$(H) curve of $Nd_3BWO_9$ at 2 K demonstrates saturation tendency up to 14 T with $M_S$ = 1.28 $\mu_B$/$Nd^{3+}$ (Figure 5b). This value is close to the half value of low temperature fitted moment as observed in other $Nd^{3+}$ based oxides.[28,29]

**$Gd_3BWO_9$.** The $\chi(T)$ of $Gd_3BWO_9$ is shown in Figure 4c, no signature of magnetic order is seen down to 2 K. The high-temperature Curie-Weiss fitting of $\chi^{-1}(T)$ yields $\theta_{CW}$ = -3.64 K and $\mu_{eff}$ = 8.6 $\mu_B$. This moment calculated here is slightly larger than theoretical value, $g_J[J(J+1)]^{1/2}$ = 7.94 $\mu_B$/$Gd^{3+}$ known for the $Gd^{3+}$ ions having $^8S_{7/2}$ multiplet with $g$ =2. The low-temperature fitting between gives $\theta_{CW}$ = -0.89 K and $\mu_{eff}$ = 8.52 $\mu_B$/$Gd^{3+}$, where the negative $\theta_{CW}$ value reflects the dominant AFM coupling between $Gd^{3+}$ ions persistent down to low temperatures. Since the $Gd^{3+}$ ($^4f_7$, S= 7/2, L= 0) ions have not orbital moment as 1$^{st}$ order approximation, $Gd^{3+}$ moments usually show quasi-isotropic magnetic interactions due to the absence of CEF effect. The field-dependent magnetization at 2 K display nonlinear field dependence and saturates above ~ 4 T (Figure 5c), reaching maximum of $M_S$ ~ 7.68 $\mu_B$ at 7 T. This value is slightly larger than $M_S$= $g_JJ\mu_B$ = 7 $\mu_B$/$Gd^{3+}$ but similar to the value ~ 7.6$\mu_B$/$Gd^{3+}$ and ~7.24$\mu_B$/$Gd^{3+}$ for $GdZnAl_{11}O_{19}$ [28] and $KBaR(BO_3)_2$, [30] respectively. Here, it is noted that the $M$(H) data at different temperatures are fitted-well with Brillouin function, suggestive of the PM-type coupled nature as Heisenberg-like spins of $Gd^{3+}$ ions.

**$Tb_3BWO_9$.** The $\chi$(T) of $Tb_3BWO_9$ (Figure 4d) shows PM behaviors in all measured temperature regions. High temperature Curie-Weiss analysis of $\chi^{-1}(T)$ provides $\theta_{CW}$= -3.24 K and $\mu_{eff}$ = 10.03$\mu_B$/$Tb^{3+}$, which is fairly close to the value of 9.72 $\mu_B$ expected for the free $Tb^{3+}$ ions (4$f^8$, $^7F_6$) with $m_J$ = ±6 doublet and $g$= 3/2. At low-temperature region (8 K - 25 K), similar analysis results in $\theta_{CW}$ = -0.79 K and $\mu_{eff}$ = 9.61 $\mu_B$/$Tb^{3+}$. The $M(H)$ curve at 2 K show nonlinear behaviour at low fields ($H$< 4 T), above which, it shows linear field dependence due to the Van Vleck PM contributions, as shown in Figure 5d. The observed maximum $M_S$~ 4.94$\mu_B$/$Tb^{3+}$, is around half of effective moment indicative of Ising-type anisotropy.

**$Dy_3BWO_9$.** The inverse $\chi^{-1}(T)$ for $Dy_3BWO_9$ exhibits PM behavior displayed in Figure 4e. Linear fitting of $\chi^{-1}(T)$ at high temperatures gives $\theta_{CW}$ = -5.81 K and $\mu_{eff}$ = 10.86 $\mu_B$/$Dy^{3+}$, while a similar fit between 8 K and 25 K leads to $\theta_{CW}$= -0.82 K and $\mu_{eff}$ = 10.49 $\mu_B$/$Dy^{3+}$. The observed effective moment is in agreement with 10.63 $\mu_B$/$Dy^{3+}$ of free $Dy^{3+}$($^6H_{15/2}$) ion. As shown in Figure 5e, the magnetization at 2 K increases rapidly below 2 T, and gradually saturates at ~ 3 T. The maximum magnetization ~ 5.18 $\mu_B$/$Dy^{3+}$ is slightly larger than half of full polarized $Dy^{3+}$ moments with $g_JJ\mu_B$ = 10 $\mu_B$/$Dy^{3+}$.

**$Ho_3BWO_9$.** As depicted in Figure 4f, low-temperature linear fitting of $Ho_3BWO_9$ yields $\theta_{CW}$ = -1.14 K and $\mu_{eff}$ = 10.38 $\mu_B$/$Ho^{3+}$, close to the value of 10.60 $\mu_B$/$Ho^{3+}$ for free $Ho^{3+}$ ($^5I_8$) ions. The magnetizations at different temperature up to 7 T are shown in Figure 5f. The $M(H)$ curves at 2 K show nonlinear field response and saturates at ~4 T, with maximum value $M_S$= 5.18 $\mu_B$/$Ho^{3+}$ at 7 T.

This value is in good agreement with half of $M_S = g_J J \mu_B = 10$ $\mu_B/Ho^{3+}$, due to the powder averaging of Ising anisotropic spins.

For $RE^{3+}$ based oxides, the crystal field of $RE^{3+}$ ions with a total ground state angular $J$ momentum splits into $2J+1$ multiplets that determine the magnetic properties of single ions. The RE ions containing odd number of $4f$ electrons like $Nd^{3+}$, $Gd^{3+}$, and $Dy^{3+}$ are Kramer ions, where the time-reversal symmetry restrictively protects the single ion ground doublet and remains degenerate. The RE ions with even numbers of $4f$ electrons such as $Pr^{3+}$, $Tb^{3+}$ and $Ho^{3+}$, the ground state doublet is non-Kramer doublet, which not necessarily be degenerated. Taking into account the different local ground-state doublet of $4f$ electrons, it is useful to perform comparative study on the magnetic ground state and magnetic anisotropy on different RE-based frustrated magnets. Additionally, in such RE-based magnetic oxides, the coordination environment of $RE^{3+}$ ions and resulting CEF can strongly affect its' magnetic properties. To more clearly clarify their magnetic behaviors, here we do comparisons on magnetic parameters of present $RE_3BWO_9$ with various RE-based frustrated magnets, including the same 8-fold coordinated $RE_3Sb_3Mg_2O_{14}$ (RE = Pr-Ho),[16] pyrochlore $RE_2Ti_2O_7$ (RE = Gd-Ho),[12,31-33] $RE_2Sn_2O_7$ (RE = Pr-Ho)[34] and $RE_2Pb_2O_7$ (R = Pr-Gd),[35] and 6-fold coordinated $KBaRE(BO_3)_2$ (RE = Gd-Ho),[30] 12-fold $REZnAl_{11}O_{19}$ (RE = Pr-Tb),[27] $NaRES_2$ (RE = Nd-Ho) systems,[35] and $Li_3RE_3Te_2O_{12}$ (RE = Pr-Ho),[36] as shown in Figure 6a,b. For all compounds, the effective moments obtained from high-temperature fit give good consistency with values of free $RE^{3+}$ ions. The low temperature fitted moments have different deviations for different systems, which can be due to different coordinated environment of $RE^{3+}$ ions and resultant CEF effect.

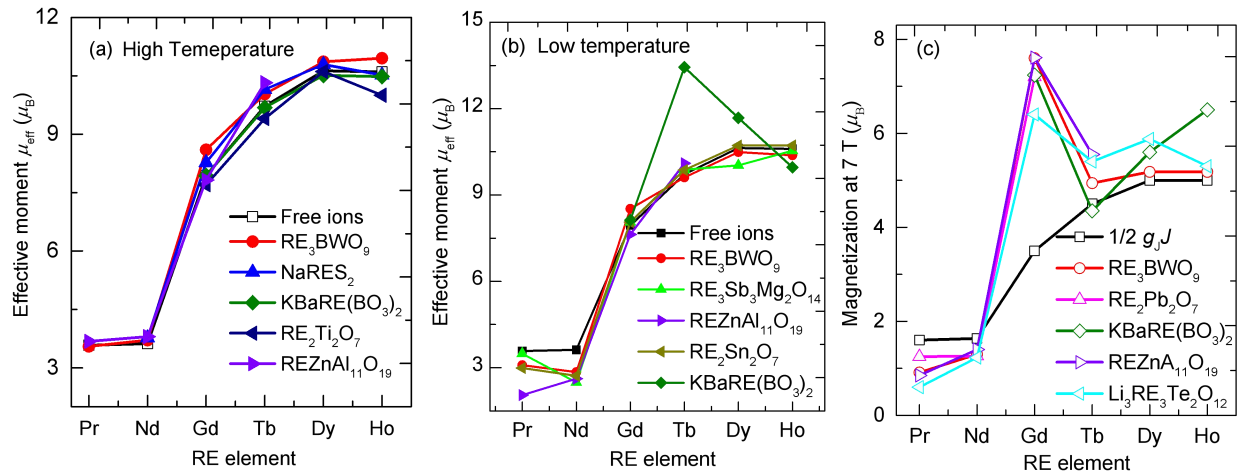

Fig. 6. Comparisons of the fitted effective magnetic moments by Curie-Weiss Law from high temperature susceptibilities (a) and low temperature susceptibilities(b), (c) the related comparison of saturated magnetizations under $H$= 7 T at 2 K for $RE_3BWO_9$, $RE_3Sb_3Mg_2O_{14}$, $REZnAl_{11}O_{19}$, $NaRES_2$, $RE_2Ti_2O_7$, $Li_3RE_3Te_2O_{12}$, $RE_2Sn_2O_7$ and $RE_2Pb_2O_7$ systems.

Figure 6c provides the comparisons of obtained maximum magnetization $M_S$ at 2 K under field of 7 T. The $M_S$ of Gd members are close to the expected value of Heisenberg-like spins of $Gd^{3+}$ ions. For other RE members, the *Ms* is nearly half-value of expected moments for $Tb^{3+}$, $Dy^{3+}$ and $Ho^{3+}$ systems indicative of Ising anisotropy. For the $Pr^{3+}$ and $Nd^{3+}$-containing compounds, the observed saturated magnetizations are close to half value of low temperature fitted effective moments, indicative of Ising-like anisotropy. Given that the presence of different spin types (Ising, Heisenberg, and planar XY) and coexistent competing exchange/dipolar interactions, different magnetic ground states are expected to be produced in $RE_3BWO_9$, just like the proposed chiral dipolar spin order in $Nd_3Mg_2Sb_3O_{14}$,[38] 120° long-range order in $Gd_3Mg_2Sb_3O_{14}$,[16] emergent charge order in $Dy_3Mg_2Sb_3O_{14}$,[13] etc.

From structure viewpoints, $RE_3BWO_9$ show some similar features with $RE_3Ga_5SiO_{14}$.[30] As example, for RE=Pr system, $Pr_3Ga_5SiO_{14}$ have nearest neighbor RE-RE distances ~4.2 Å within the Kagomé layers close to the ones of $Pr_3BWO_9$, and relatively larger interlayer distances (~ 5 Å). So, these two systems share comparable exchange energy. The dipolar energy of $Pr_3BWO_9$ is roughly $D_{nn}= \mu_o\mu_{eff}^2/4\pi(r_{nn})^3$ ~ 1.1 K and ~0.7 K by taking $\mu_{eff}$= 3.55 $\mu_B$ and nearest neighbor intralayer Pr-Pr distance $r_{nn}$ = 4.3 Å and 4.9 Å, which is comparable to the dipolar interactions ~1.2 K of $Pr_3Ga_5SiO_{14}$ system (with $\mu_{eff}$ = 3.6 $\mu_B$ and $r_{nn}$ = 4.2 Å). But, the $RE_3BWO_9$ is a 3D coupled systems due to comparable interlayer/intralayer RE-RE distances different from $RE_3Ga_5SiO_{14}$ and $RE_3Sb_3Mg_2O_{14}$, which are 2D Kagomé system with large RE-RE interlayer distances. More importantly, $RE_3BWO_9$ manifests a free of site disorder and rich spin-types, provide a promising model material for investigating exotic magnetic ground states. Further research on low-temperature thermodynamic and magnetic excitation based on large single-crystal will be substantial to clarify the magnetic state of these systems.

## ■ CONCLUSIONS

A new family of rare-earth-based Kagomé lattice antiferromagnets, $RE_3BWO_9$ (La, Pr, Nd, Gd-Ho) borotangstanates were successfully synthesized and magnetically characterized, where magnetic $RE^{3+}$ spins are located on the distorted Kagomé lattice within *ab* plane. The structural analysis reveal these compounds are free of chemically site-mixing occupancy of RE and B/W cations, avoiding the influence of exchange randomness on its magnetic behaviors. All serial $RE_3BWO_9$ compounds show dominant antiferromagnetic interactions between the $RE^{3+}$ magnetic moments without long-range magnetic order down to 2 K. The analyses of isothermal magnetizations reveal the presence of different magnetic anisotropy for different RE ions, making $RE_3BWO_9$ as potential system for investigating diverse magnetic ground states.

## ■ AUTHOR INFORMATION

**Corresponding Authors**
*E-mail: tianzhaoming@hust.edu.cn
*E-mail: syl@zzuli.edu.cn

**Notes**

The authors declare no competing financial interest.

## ■ ACKNOWLEDGEMENTS

We would like to thanks Gang Chen for helpful discussions. We acknowledge financial support from the National Natural Science Foundation of China (Grant No. 11874158 and 11604281), and Fundamental Research Funds for the Central Universities (Grant No. 2018KFYYXJJ038 and 2019KFYXKJC008). We would like to thank the staff of the analysis centre of Huazhong University of Science and Technology for their assistance in structural characterization and analysis.

## ■ REFERENCES